\documentclass[journal,onecolumn]{IEEEtran}
\UseRawInputEncoding
\usepackage[ruled,linesnumbered]{algorithm2e}

\usepackage{color}

\usepackage{subfigure,array}
\usepackage{graphicx,cite,epsfig,amssymb,amsmath,multirow,lettrine,flushend,extarrows}

\begin{document}



\title{A Review of Deep Reinforcement Learning for Smart Building Energy Management}

\author{{Liang~Yu,~\IEEEmembership{Member,~IEEE}, Shuqi Qin, Meng Zhang, Chao Shen,~\IEEEmembership{Senior Member,~IEEE},\\ Tao~Jiang,~\IEEEmembership{Fellow,~IEEE}, and Xiaohong~Guan,~\IEEEmembership{Fellow,~IEEE}}
\thanks{
\newline L. Yu is with the College of Automation \& College of Artificial Intelligence, Nanjing University of Posts and Telecommunications, Nanjing 210003, China, and also with Xi'an Jiaotong University, Xi'an 710049, China. (email: liang.yu@njupt.edu.cn) \newline
S. Qin is with the College of Internet of Things, Nanjing University of Posts and Telecommunications, Nanjing 210003, China. \newline
M. Zhang and X. Guan are with Systems Engineering Institute, Ministry of Education Key Lab for Intelligent Networks and Network Security, Xi'an Jiaotong University, Xi'an 710049, China.  \newline
C. Shen is with the School of Cyber Science and Engineering, Xi'an Jiaotong University, Xi'an 710049, China. (email: cshen@sei.xjtu.edu.cn) \newline
T. Jiang is with Wuhan National Laboratory for Optoelectronics, School of Electronic Information and Communications, Huazhong University of Science and Technology, Wuhan 430074, China. (email: Tao.Jiang@ieee.org) \newline
}}

\markboth{IEEE Internet of Things Journal,~Vol.~XX, No.~XX, Month~2021}%
{Yu \MakeLowercase{\textit{et al.}}: A Review of Deep Reinforcement Learning for Smart Building Energy Management}

\maketitle

\begin{abstract}
Global buildings account for about 30\% of the total energy consumption and carbon emission, raising severe energy and environmental concerns. Therefore, it is significant and urgent to develop novel smart building energy management (SBEM) technologies for the advance of energy-efficient and green buildings. However, it is a nontrivial task due to the following challenges. Firstly, it is generally difficult to develop an explicit building thermal dynamics model that is both accurate and efficient enough for building control. Secondly, there are many uncertain system parameters (e.g., renewable generation output, outdoor temperature, and the number of occupants). Thirdly, there are many spatially and temporally coupled operational constraints. Fourthly, building energy optimization problems can not be solved in real-time by traditional methods when they have extremely large solution spaces. Fifthly, traditional building energy management methods have respective applicable premises, which means that they have low versatility when confronted with varying building environments. With the rapid development of Internet of Things technology and computation capability, artificial intelligence technology find its significant competence in control and optimization. As a general artificial intelligence technology, deep reinforcement learning (DRL) is promising to address the above challenges. Notably, the recent years have seen the surge of DRL for SBEM. However, there lacks a systematic overview of different DRL methods for SBEM. To fill the gap, this paper provides a comprehensive review of DRL for SBEM from the perspective of system scale. In particular, we identify the existing unresolved issues and point out possible future research directions.
\end{abstract}


\section{Introduction}\label{s1}
Buildings account for a large portion of total energy consumption and total carbon emission in the world\cite{Hu2019,Hu2018,Park2019,Dong2020,Haghi2019}. For example, global buildings consumed 30\% of total energy and generated 28\% of total carbon emission in 2019\cite{Global2020}. Moreover, the energy demand of buildings is expected to increase by 50\% in the next 30 years\cite{Sharma2019}\cite{Zhou2018}. Under the above background, smart buildings have received more and more attention in recent years, which can provide sustainable, economical, and comfortable operational environments for occupants using many advanced technologies, e.g., Internet of Things (IoT), cloud computing, machine learning, and big data analytics\cite{QOLOMANY2019,Minoli2017,Zhangxiangyu2017}. For supporting the above features, it is significant and urgent to develop novel smart building energy management (SBEM) technologies\cite{Feng2020}, which can implement the optimal tradeoff among energy consumption, carbon emission, energy cost, and user comfort\cite{Yangbin2020,WangF2018,Pallante2020,Ahmad2020,ZhangRufeng2020,Gao2020} by intelligently scheduling building energy systems.

Although SBEM has many advantages, the following challenges have to be addressed. Firstly, due to the existence of many complex and random factors, it is often intractable to develop an explicit building thermal dynamics model that is accurate and efficient enough for building energy optimization\cite{Wei2017}. Secondly, there are many uncertain system parameters\cite{YuIoT2019}, e.g., renewable generation output, electricity price, indoor temperature, outdoor temperature, $\text{CO}_2$ concentration, and the number of occupants. Thirdly, there are many temporally and spatially coupled operational constraints related to energy subsystems\cite{YuTSG2021}\cite{ZhangChi2019}, e.g., heating, ventilation, and air conditioning (HVAC) systems, and energy storage systems (ESSs), which means that the current system decision will affect the future decisions and the decisions among different subsystems should be coordinated. Fourthly, it is difficult to solve large-scale building energy optimization problems in real-time when traditional optimization methods are adopted\cite{Mocanu2019}. To be specific, any time when an optimization is needed, these methods have to compute completely or partially all the possible solutions and choose the best one. When the solution space is very large, the computation process is time-consuming\cite{YuTSG2021}. Finally, it is hard to develop a generalized building energy management method that can be applied in all building environments\cite{Gao2020}. In existing SBEM methods, most of them have strong applicable premises\cite{Zhangzhidong2020}, e.g., stochastic programming and model predictive control (MPC) need the prior or forecasting information of uncertain parameters\cite{Ma2015}\cite{Gaun2010TSG}, and Lyapunov optimization techniques require some strict usage conditions\cite{Ahmad2020}\cite{Yu2018JIOT}.

\begin{table*}[htbp]
\center
\caption{The Comparison between Our Work and Related Surveys}\label{table_1} \centering
\begin{tabular}{|m{2.5cm}<{\centering}|m{2.4cm}<{\centering}|m{2.4cm}<{\centering}|m{3.2cm}<{\centering}|m{1.8cm}<{\centering}|m{2.2cm}<{\centering}|}
\hline
\textbf{Literature}& \textbf{Main focus} & \textbf{System type(s)} & \textbf{Involved Methods/Algorithms}& \textbf{DRL methods for SBEM classified}& \textbf{Future directions in DRL-based SBEM provided}\\
\hline
\hline
Han \emph{et al.}\cite{Han2019}&  Occupant comfort control & HVAC, lighting systems   & RL & No &	No\\
\hline
Leit$\tilde{a}$o \emph{et al.} \cite{LEIT2020} & Building energy optimization & Smart home  & LP, NLP, CP, DP, GA, PSO, MPC, RL & No & No \\
\hline
Mason \emph{et al.}\cite{Mason2019}&  Building energy optimization & HVAC, EWH, home management systems, smart home  & RL & No &	No\\
\hline
Wang \emph{et al.}\cite{WangZ2020}&  Building energy optimization & HVAC, batteries, home appliances, EWH, windows, lighting   & RL & No &	No\\
\hline
Rajasekhar \emph{et al.}\cite{Rajasekhar2020}& Building energy optimization & HVAC  & RNN, WNN, RT, SVM, PSO, MPC, FL, RL, DQN  & No &	 No\\
\hline
Zhang \emph{et al.}\cite{Zhangdongxia2018} & Cyber security, demand response, load forecasting, and microgrid & Smart grid  & RL, DQN, DDPG, NAF, A3C & No & No \\
\hline
Yang \emph{et al.}\cite{Yang2020}& Energy and electric system security, operation optimization & Microgrid, ESS, HVAC, home appliances, PV   & RL, DQN, DDPG, A3C, DDQN, TRPO & No &	No\\
\hline
Our work & Building energy optimization & A single building energy subsystem, multiple energy subsystems in buildings, building microgrids &  DQN, DDQN, BDQ, DDPG, PDDPG, MADDPG, FH-DDPG, A2C, A3C, TRPO, PPO, MAPPO, MuZero, MAAC, EB-C-A2C, EB-C-DQN  & Yes &  Yes\\
\hline
\end{tabular}
\end{table*}

As a general artificial intelligence technology, deep reinforcement learning (DRL)\cite{Mnih2015}\cite{YuxiLi2017} is promising to address the above challenges and has been applied in many fields, e.g., games\cite{Schrittwieser2020}\cite{Shao2019}, autonomous driving\cite{ChenJ2019,Kiran2021,Aradi2021,Haydari2021}, autonomous IoT\cite{Lei2019}, smart buildings\cite{Wei2017}\cite{YuIoT2019}\cite{Lee2020}\cite{Zhangxiangyu2020}, smart city\cite{Mohammadi2018}, wireless networks\cite{Sun2019}, Internet of energy\cite{LinL2020}, unmanned aerial vehicles\cite{Wangc2020}, smart microgrids\cite{LeiTan2020}, edge computing\cite{Chenxianfu2019}, and manufacturing systems\cite{Lu2020}. In 2017, the first work that adopts DRL algorithm for SBEM has been done\cite{Wei2017}. To be specific, a deep Q-network (DQN) algorithm has been adopted for the control of building HVAC systems and simulation results have showed the effectiveness of the designed control algorithm in reducing energy cost and maintaining thermal comfort of occupants. Since then, many DRL-based methods for SBEM have been proposed\cite{YuIoT2019}\cite{YuTSG2021}\cite{Lork2020}\cite{Hu2020}. In general, DRL-based methods have following advantages in dealing with the above-mentioned five challenges:

\begin{itemize}
  \item \textbf{For challenge~1}: Based on the information interacted with actual building environments, DRL agents can learn the optimal control policies by trial and error. Therefore, DRL can support system operation without knowing explicit building thermal dynamics models\cite{Wei2017}.
  \item \textbf{For challenge~2}: After the training process is finished, the trained DRL agent will be used for performance testing. Given the current state of an actual environment, the DRL agent will generate an action via a mapping function. Since no forecasting or statistics information of building environments is used in the above process, DRL can tackle system uncertainties\cite{YuIoT2019}\cite{HuYue2018}.
  \item \textbf{For challenge~3}: By designing proper reward functions, building energy subsystems can coordinate with each other under the framework of multi-agent DRL. As a result, spatially-coupled operational constraints are guaranteed\cite{YuTSG2021}. Moreover, by choosing reasonable actions or designing efficient reward functions, temporally-coupled operational constraints related to energy subsystems (e.g., HVAC systems and ESSs) can be satisfied\cite{YuIoT2019}.
  \item \textbf{For challenge~4}: During the testing phase, the computational complexity of the DRL algorithm is very low since just the forward propagation in deep neural networks (DNNs) is involved. Even if a high-dimensional raw state is given, the optimal control actions can be determined instantly (e.g., few milliseconds)\cite{Mocanu2019}\cite{Nagarathinam2020}\cite{Yang2019}.
  \item \textbf{For challenge~5}: Since simulated or real data is used for training agents, the applications of DRL methods do not require rigorous mathematical models and premise conditions. Moreover, the trained DRL agent can still work or even be improved persistently by online learning when confronted with varying building environments\cite{Mocanu2019}\cite{Nagy2018}. Thus, DRL methods have wide applicable premises in solving SBEM problems.
\end{itemize}

There are many surveys related to DRL in the literature, e.g., the applications of DRL in power and electric systems, communications and networking, autonomous driving, autonomous IoT, cyber security, and multi-agent systems can be found in \cite{Zhangzhidong2020,Kiran2021,Aradi2021,Haydari2021,Luong2019,Lei2019,NguyenCS2020,NguyenMAS2019,Zhangdongxia2018,Yang2020}. However, they do not consider DRL for SBEM. Although there are several surveys on building energy systems, the involved methods are RL\cite{Han2019,LEIT2020,Mason2019,WangZ2020} or other artificial intelligence methods (e.g., MPC and fuzzy logic (FL))\cite{Rajasekhar2020}. Based on the above observation, we are motivated to conduct a comprehensive review on DRL for SBEM. For convenience, we provide the comparison between our work and related surveys in Table~\ref{table_1}. It can be observed that our work mainly focuses on DRL for SBEM from the perspective of different building system scales (i.e., a single building energy subsystem, multiple energy subsystems in buildings, and building microgrids). Moreover, we provide a systematic overview of different DRL methods for SBEM. Above all, we identify the existing unresolved issues and point out possible future directions. We hope that this paper can show some insights in this direction and raise the attention of SBEM research community to explore and exploit DRL, as another alternative or even a better solution for SBEM.

The rest of this paper is organized as follows. In Section \ref{s2}, we give an overview of DRL. In Section~\ref{s3}, we introduce the background of SBEM, the procedure of solving SBEM problems using DRL, and the classification of DRL methods for SBEM. In next three sections, we discuss DRL applications in a single building energy subsystem, multiple energy subsystems of buildings, and building microgrids. In Section \ref{s7}, we identify some unsolved issues and point out the future research directions. Finally, conclusions and lessons learned are provided in Section \ref{s8}. For easy understanding, the list of abbreviations in alphabetical order is provided in Table~\ref{table_2}.

\begin{table}[htbp]
\center
\caption{List of abbreviations in alphabetical order}\label{table_2} \centering
\begin{tabular}{|c|c|}
\hline
\textbf{Abbreviation} &\textbf{Description}\\
\hline
A2C &Advantage Actor-Critic\\
\hline
A3C  &Asynchronous Advantage Actor-Critic\\
\hline
AHU &Air Handling Unit\\
\hline
BAS  &Building Automation System\\
\hline
BDQ &Branching Dueling Q-Network\\
\hline
BEM &Building Energy Model\\
\hline
CD  &Clothes Dryer\\
\hline
CNN/DNN &Convolutional/Deep Neural Network\\
\hline
CP  &Convex Programming\\
\hline
D-DNFQI  & Double Deep Neural Fitted Q Iteration\\
\hline
DDPG  &Deep Deterministic Policy Gradient\\
\hline
DDQN   &Double Deep Q-Network\\
\hline
DG  &Diesel Generator\\
\hline
DP  &Dynamic Programming\\
\hline
DQN &Deep Q-Network\\
\hline
DRL  &Deep Reinforcement Learning\\
\hline
DW  &Dishwasher\\
\hline
EB-C-A2C &Entropy-Based Collective Advantage Actor-Critic\\
\hline
EB-C-DQN  &Entropy-Based Collective Deep Q-Network\\
\hline
EHP  &Electric Heat Pump\\
\hline
ESS  &Energy Storage System\\
\hline
EV  &Electric Vehicle\\
\hline
EWH   &Electric Water Heater\\
\hline
FH-DDPG &Finite-Horizon Deep Deterministic Policy Gradient\\
\hline
FH-RDPG &Finite-Horizon Recurrent Deterministic Policy Gradient\\
\hline
FL  &Fuzzy Logic\\
\hline
GA  &Genetic Algorithm\\
\hline
GB  &Gas Boiler\\
\hline
HVAC  &Heating, Ventilation and Air Conditioning\\
\hline
IoT  &Internet of Things \\
\hline
LP  &Linear Programming\\
\hline
LSTM  &Long Short-Term Memory\\
\hline
MAAC   &Multi-Actor Attention-Critic\\
\hline
MADDPG &Multi-Agent Deep Deterministic Policy Gradient\\
\hline
MAPPO  &Multi-Agent Proximal Policy Optimization\\
\hline
MCTS  &Monte-Carlo Tree Search\\
\hline
MDP  &Markov Decision Process\\
\hline
MPC &Model Predictive Control\\
\hline
NAF   &Normalized Advantage Functions\\
\hline
NLP &Non-Linear Programming\\
\hline
PDDPG  &Prioritized Deep Deterministic Policy Gradient\\
\hline
PILCO &Probabilistic Inference for Learning COntrol\\
\hline
PPO &Proximal Policy Optimization\\
\hline
PSO  &Particle Swarm Optimization\\
\hline
PV  &Photovoltaic\\
\hline
RL &Reinforcement Learning \\
\hline
RNN &Recurrent Neural Network\\
\hline
RT  &Regression Tree\\
\hline
SBEM  &Smart Building Energy Management\\
\hline
SVM &Support Vector Machine\\
\hline
TES  &Thermal Energy Storage\\
\hline
TRPO  &Trust Region Policy Optimization\\
\hline
VAV  &Variable Air Volume\\
\hline
WM &Washing Machine\\
\hline
WNN  &Wavelet Neural Network\\
\hline
WT  &Wind Turbine \\
\hline
\end{tabular}
\end{table}

\section{An Overview of Deep Reinforcement Learning}\label{s2}

\begin{figure}[!htb]
\centering
\includegraphics[scale=0.29]{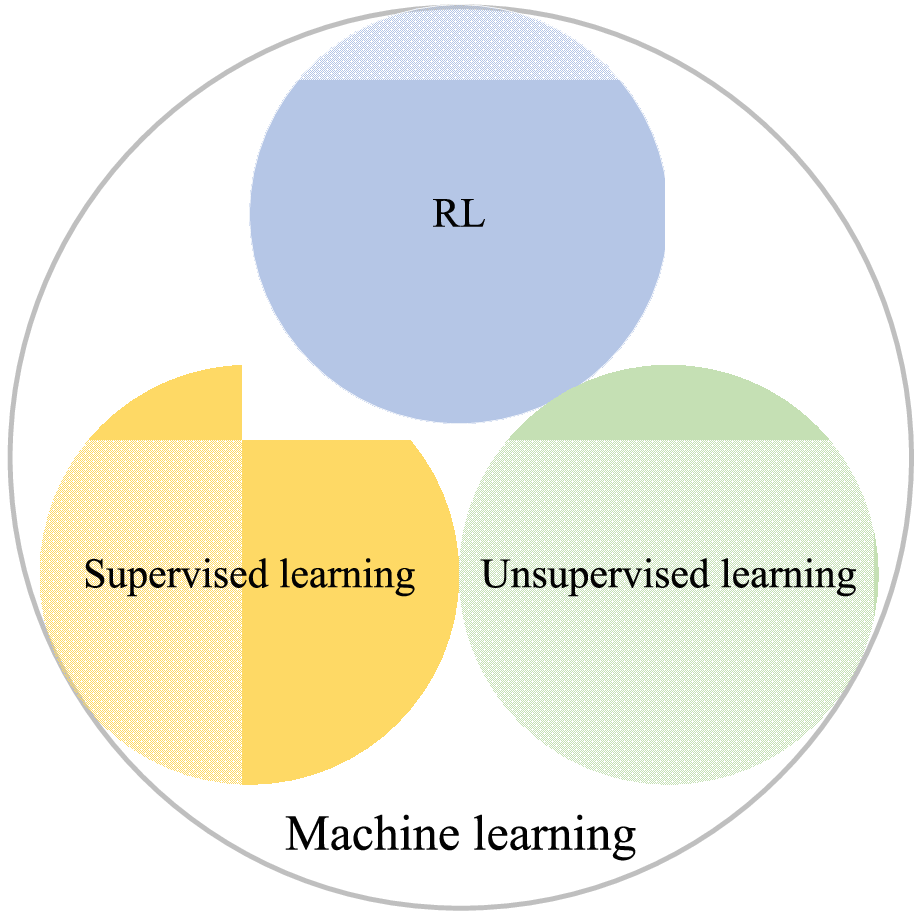}
\caption{Classification of machine learning}\label{fig_1}
\end{figure}

According to the ways of feedback, machine learning can be divided into three types as shown in Fig.~\ref{fig_1}, i.e., supervised learning, unsupervised learning, and reinforcement learning (RL)\cite{Sutton2018}. As for supervised learning, an immediate feedback can be obtained by comparing the prediction value with the real value that given by the label data, which will be used to improve predictor. In contrast, no feedback can be received in unsupervised learning since the input data is not labeled. While interacting with an environment, a delayed feedback is involved in RL since the action taken at the current state will affect future states and actions, and the value of taking an action at the current state can not be known immediately but be learned gradually. Typically, supervised learning and unsupervised learning are used to solve single-stage problems (e.g., regression, classification, clustering, and dimension reduction), but RL is specialized in solving multi-stage decision problems\cite{Sutton2018}. Under the background of SBEM, supervised learning can be used to develop building thermal dynamics models and reward models. Based on these models, RL can reduce the number of interactions with the environment, resulting in a high sampling efficiency.

\begin{figure*}[!htb]
\centering
\includegraphics[scale=0.94]{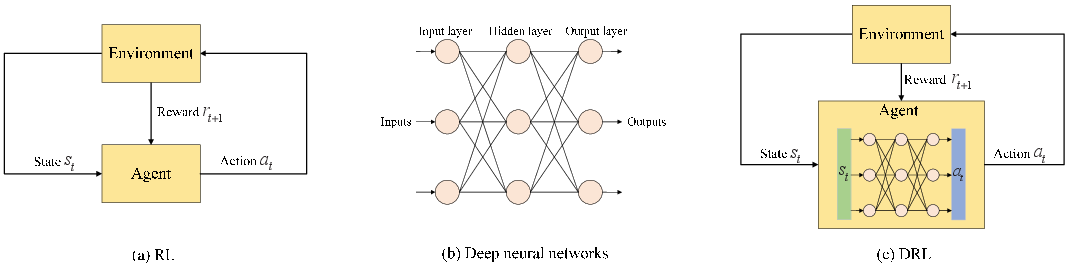}
\caption{Illustration of RL, deep learning, and DRL}\label{fig_2}
\end{figure*}

 DRL can be regarded as the combination of deep learning and RL as shown in Fig.~\ref{fig_2}. To be specific, DNNs are used to approximate the optimal value functions or optimal policies in RL. Therefore, DRL has a powerful representation ability and strong decision-making ability under uncertainty\cite{Wangzhe2019}\cite{Raman2020}. Since DRL algorithms are mainly based on Markov decision process (MDP) framework or its variants (e.g., partially observable MDP\cite{Lei2019} and Markov game\cite{YuTSG2021}), we first give the background of MDP. Then, we introduce some terms (e.g., \emph{policy}, \emph{action-value function}, \emph{experience replay}, and \emph{target network}) in RL and DRL, which will be mentioned frequently in next several sections.

\subsection{MDP}
Typically, an MDP is defined by a five-tuple ($\mathcal{S}$, $\mathcal{A}$, $P$, $\mathcal{R}$, $\gamma$), where $\mathcal{S}$ and $\mathcal{A}$ denote the sets of state and action, respectively. $P:~\mathcal{S}\times \mathcal{A} \times \mathcal{S}\rightarrow [0,1]$ denotes the state-transition probability function $P(s'|s,a)$ ($s',s\in \mathcal{S}$, $a\in \mathcal{A}$), which models the uncertainty in the evolution of system states based on the action taken by the agent. $\mathcal{R}:\mathcal{S}\times \mathcal{A}\rightarrow \mathbb{R}$ is the reward function and $\gamma\in [0,~1]$ is a discount factor.
Note that MDP provides a mathematical framework for multi-stage optimal decision problems under uncertainty. In other words, the decision maker (i.e., the agent) observes a state $s_{t}$ and takes an action $a_{t}$ at each time slot $t$. Next, the state of the system (i.e., the environment) evolves into another one. Then, the agent finds itself in a new state $s_{t+1}$ and receives a reward $r_{t+1}$. In addition, the aim of the agent at time slot $t$ is to maximize the expected return it receives over the future\cite{Sutton2018}, which is given by $\sum\nolimits_{k=0}^{\infty}\gamma^k r_{t+k+1}$.

\subsection{RL}
RL has been widely used in solving MDPs\cite{Sutton2018,Deng2020,Canteli2019,Han2019,KimS2018,Ahrarinouri2021,LuTSG2019,Ruelens2017}. In an RL process, the agent learns its optimal policy $\pi$ by interacting with the environment, where a \emph{policy} $\pi$ is a mapping from states to the probabilities of selecting every possible action\cite{Sutton2018}. In particular, the agent observes a state and takes an action at slot $t$. Then, it receives a reward and a new state, which are used to update the policy. The above process repeats until the policy converges.

To better illustrate the key idea of RL, Q-learning is introduced in this subsection, which is one of the most classic RL algorithms that learn a deterministic policy indirectly. In other words, Q-function (i.e., \emph{action-value function}) is learned for selecting decisions instead of policy function itself. Let the value of taking action $a$ in state $s$ under a policy $\pi$ be $Q_{\pi}(s,a)$, which is defined by
\begin{align} \label{f_1}
Q_{\pi}(s,a)\doteq \mathbb{E}_{\pi}[\sum\nolimits_{k=0}^{\infty}\gamma^k r_{t+k+1}(s_{t}=s,a_{t}=a)],
\end{align}
where $\mathbb{E}_{\pi}[\cdot]$ denotes the expected value of a random variable given that the agent follows policy $\pi$. Then, the optimal action-value function $Q^*(s,a)$ is $\max_{\pi}Q_{\pi}(s,a)$ and can be calculated by the Bellman optimality equation in a recursive manner\cite{YuIoT2019} as follows,
\begin{equation}\label{f_2}
\begin{array}{l}
{Q^*}(s,a) =\mathbb{E}[r_{t+1} + \gamma {\max _{{a}'}}{Q^*}({s_{t + 1}},{a}')|{s_t} = s,{a_t} = a]\\ \nonumber
\quad \quad \quad \quad  = \sum\nolimits_{s',r} P (s',r|s,a)[r+ \gamma {\max _{{a}'}}{Q^*}(s',{a'})],
\end{array}
\end{equation}
where $s'\in \mathcal{S}$, $r\in \mathcal{R}$, $a'\in \mathcal{A}$, and $P(s',r|s,a)$ denotes a conditional probability function. To obtain the value of ${Q^*}(s,a)$, the information of $P(s',r|s,a)$ must be known, which may be unavailable in practice. To address this challenge, Q-learning algorithm is proposed to approximate ${Q^*}(s,a)$ as follows,
\begin{align} \label{f_3}
Q(s_{t},a_{t})\leftarrow Q(s_{t},a_{t})+\Delta_t,
\end{align}
where $\Delta_t=\alpha \Big[r_{t+1}+\gamma\max_{a'}Q(s_{t+1},a')-Q(s_{t},a_{t})\Big]$ and $\alpha$ is the step size. It is obvious that $Q(s_{t},a_{t})=r_{t+1}+\gamma\max_{a'}Q(s_{t+1},a')$ when $\Delta_t=0$. At this time, $Q(s_{t},a_{t})$ will not be updated and the learned action-value function $Q$ directly approximates the optimal action-value function ${Q^*}(s,a)$. Note that Q-learning algorithm is effective when state space is low-dimensional. To support high-dimensional state space, a nonlinear function approximator such as a neural network can be used to represent the action-value function in RL. At this time, RL is known to be unstable or even divergent.

\subsection{DRL}
As the first DRL algorithm, DQN can overcome the above-mentioned drawback of Q-learning by adopting several techniques of stabilizing learning process, e.g., \emph{experience replay} and \emph{target network}\cite{Mnih2015}. To be specific, experience replay mechanism stores the experience transitions $(s_{t},a_{t},s_{t+1},r_{t+1})$ in a replay memory and draw samples of them uniformly at random for training, which brings greater data efficiency when compared with standard online Q-learning algorithm. Moreover, randomizing the samples contributes to the reductions of their correlations and the variance of updating DNN weights. In addition, target network is adopted to improve the stability of training process by copying a separate network with longer update period for the computation of target value (i.e., $r_{t+1}+\gamma\max_{a'}Q(s_{t+1},a')$).

In addition to DQN, many DRL algorithms have been proposed in existing works. Generally speaking, these DRL algorithms can be classified into two types, i.e., model-free DRL algorithms and model-based DRL algorithms. In particular, model-free DRL algorithms do not need to know environment models (i.e., state-transition probability function $P(s'|s,a)$ or $P(s',r|s,a)$) since they learn policies based on the information directly interacted with unknown environments. Different from model-free DRL algorithms, model-based DRL algorithms need to construct environment models. According to the way of learning a policy, model-free DRL methods can be further divided into value-based methods and policy-based methods. To be specific, the former learns an approximation of optimal value function (i.e., learn a deterministic policy indirectly), while the latter learns an approximation of optimal policy directly. Typically, value-based methods sometimes update value function in an ``off-policy" (i.e., the policy to be evaluated and improved is unrelated to the policy used for sampling an action at the next state) manner, which means that the previous collected experience transitions in the same environment can be used for training and a high data efficiency can be achieved. In contrast, ``on-policy" means that all of the updates are made using the data from the trajectory distribution induced by the current policy\cite{Gu2017}. Therefore, ``on-policy" methods are more stable but less data-efficient compared with ``off-policy" methods. In addition, according to the number of agents, DRL algorithms can be divided into two types, i.e., single-agent and multi-agent DRL algorithms. In Table~\ref{table_3}, typical DRL algorithms and their categories are summarized.

\begin{table}[htbp]
\center
\caption{Classification of Typical DRL algorithms}\label{table_3} \centering
\begin{tabular}{|c|c|c|}
\hline
\textbf{DRL Algorithms} &\textbf{Category} & \textbf{Source}\\
\hline
\hline
DQN  &Value-based, off-policy, single-agent & \cite{Mnih2015}\\
\hline
Double DQN  &Value-based, off-policy, single-agent & \cite{Hasselt2016} \\
\hline
Dueling DQN  &Value-based, off-policy, single-agent & \cite{WangZ2016} \\
\hline
Prioritized DQN  &Value-based, off-policy, single-agent & \cite{Schaul2016} \\
\hline
Distributional DQN &Value-based, off-policy, single-agent & \cite{Bellemare2017}\\
\hline
Noisy DQN &Value-based, off-policy, single-agent & \cite{Fortunato2018}\\
\hline
Rainbow  &Value-based, off-policy, single-agent & \cite{Hessel2018}\\
\hline
DDPG  &Policy-based, off-policy, single-agent &  \cite{Lillicrap2016}\\
\hline
MADDPG  &Policy-based, off-policy, multi-agent & \cite{Lowe2017} \\
\hline
MAAC &Policy-based, off-policy, multi-agent & \cite{Iqbal2019} \\
\hline
Soft Actor-Critic  &Policy-based, off-policy, single-agent & \cite{Haarnoja2018}\\
\hline
A3C  &Policy-based, on-policy, single-agent & \cite{Mnih2016} \\
\hline
TRPO  &Policy-based, on-policy, single-agent &  \cite{Schulman2015} \\
\hline
PPO  &Policy-based, on-policy, single-agent &\cite{Schulman2017} \\
\hline
MAPPO  &Policy-based, on-policy, multi-agent & \cite{YuC2021} \\
\hline
Deep PILCO &Model-based, on-policy, single-agent & \cite{Gal2016} \\
\hline
World Model &Model-based, on-policy, single-agent & \cite{Ha2018} \\
\hline
MuZero  &Model-based, off-policy, single-agent & \cite{Schrittwieser2020}\\
\hline
\end{tabular}
\end{table}

\section{DRL-based Smart Building Energy Management}\label{s3}
In this section, we briefly introduce the main research problems in the field of SBEM. Then, we classify the representative DRL algorithms for SBEM by pointing out their respective advantages, disadvantages, and application scenarios, which contributes to the selection of appropriate DRL algorithms for SBEM.

\subsection{SBEM}
In smart buildings, there are several types of energy equipments, e.g., photovoltaic panels (PVs), wind turbines (WTs), diesel generators (DGs), electric energy storage systems, thermal energy storage systems, HVAC systems, lighting systems, blind systems, window systems, electric water heaters (EWHs), electric vehicles (EVs), washing machines (WMs), gas boilers (GBs), and clothes dryers (CDs). Since the operations of such equipments have considerable economic, environmental, and social impacts on buildings, it is very necessary to schedule them coordinately.
%

Considering that HVAC systems have high power consumption and can be adjusted flexibly without sacrificing user comfort, they are taken as an example to illustrate a typical research problem in SBEM field. When considering economic and social impacts, a comfort-aware energy cost minimization problem related to an HVAC system in a $N$-zone commercial building can be formulated by \textbf{P1} as follows\cite{YuTSG2021},
\begin{subequations}\label{f_4}
\begin{align}
(\textbf{P1})~&\min_{m_{i,t},\sigma_t}~\mathop \sum\nolimits_{t=1}^{L} \mathbb{E}\{C_t(m_{i,t},\sigma_t)\}  \\
s.t.&~T_i^{\min}\leq T_{i,t}\leq T_i^{\max},~\forall~i,t,~K_{i,t}>0,\\
&T_{i,t+1}=\mathcal{F}(T_{i,t},~T_{z,t}|_{\forall z\in \mathcal{N}_i},~T_{t}^{\text{out}},~m_{i,t},\varsigma_{i,t}),\\
&O_{i,t} \leq O_i^{\max},~\forall~i,t,~K_{i,t}>0, \\
&O_{i,t+1}=\mathcal{G}(O_{j,t}|_{\forall j\in \mathcal{N}},K_{i,t},m_{i,t},\sigma_t),\\
&m_{i,t} \in \mathcal{M},~\forall~i,t, \\
&\sigma_t \in \Omega,~\forall~t,
\end{align}
\end{subequations}
where $\mathbb{E}$ denotes the expectation operator, which acts on random system parameters, e.g., outdoor temperature $T_{t}^{\text{out}}$, number of occupants $K_{i,t}$. Decision variables of \textbf{P1} are air supply rate in each zone $m_{i,t}$ and damper position in air handling unit (AHU) $\sigma_t$, $L$ is the considered total number of time slots. $C_t(m_{i,t},\sigma_t)$ is the energy cost at slot $t$, $T_{i,t}$ and $O_{i,t}$ are indoor air temperature and indoor $\text{CO}_2$ concentration at slot $t$, respectively. It is obvious that they should be controlled within comfortable ranges, which can be captured by (3b) and (3d), respectively. The dynamics of $T_{i,t}$ and $O_{i,t}$ are represented by (3c) and (3e), respectively. Note that $\varsigma_{i,t}$, $\mathcal{N}_i$, and $\mathcal{N}$ are thermal disturbance, the set of neighbors related to zone $i$ ($1\leq i\leq N$), and the set of zones, respectively. The discrete solution spaces of $m_{i,t}$ and $\sigma_t$ are shown in (3f) and (3g), respectively.

To solve SBEM problem \textbf{P1} efficiently, several challenges mentioned in Section \ref{s1} have to be addressed. In addition, non-convexity and non-separability of the objective function increase the difficulty of solving \textbf{P1}. When taking all challenges into consideration, existing building energy optimization approaches are not applicable. Note that the above example is just related to the management of a single building energy subsystem. With the increase of system scale, more and more challenges are involved, which will be discussed in next four sections.

\begin{figure*}[!bht]
\centering
\includegraphics[scale=0.94]{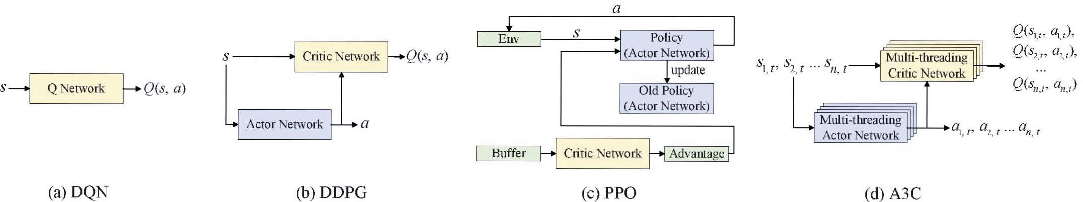}
\caption{Illustration of network architectures of four representative model-free DRL algorithms}\label{fig_3}
\end{figure*}

\subsection{Procedure of Solving SBEM Problems using DRL}
To solve SBEM problems using DRL methods, several steps can be taken as follows:

\textbf{Step~1}: Reformulating the original problem (e.g., \textbf{P1}) as an MDP problem or a Markov game (i.e., a multi-agent extension of MDP). Take \textbf{P1} for example, $N+1$ agents are adopted for the purpose of scalability, since there are $N+1$ decision variables and the solution space grows rapidly with the increase of zone number. Thus, \textbf{P1} should be formulated as a Markov game and its components (e.g., state, action, and reward function) should be designed.

\textbf{Step~2}: Designing an appropriate DRL-based algorithm for the reformulated problem. For instance, in order to solve the Markov game related to \textbf{P1}, an HVAC control algorithm has been proposed in \cite{YuTSG2021} based on multi-agent DRL with attention mechanism, which is scalable to the number of agents.

\textbf{Step~3}: Analyzing the computational complexity of the designed DRL-based algorithm in the training and testing phases. Note that a trained DRL algorithm has very low computational complexity in the testing phase since just forward propagation in DNNs is involved, e.g., DRL agent can make real-time decisions within several milliseconds given a high-dimensional environment state. As a result, computational complexity analysis of the designed DRL-based algorithm in the training phase is the priority. Typical factors that affect computational complexity of a DRL-based energy management algorithm in the training phase are summarized as follows\cite{YuIoT2019}\cite{Chung2020}, e.g., the number of hidden layers, the number of neurons in each hidden layer, the number of training episodes needed for algorithm convergence, the frequency of updating weights, and batch size.

\textbf{Step~4}: Evaluating the performances of the designed DRL-based algorithm from different perspectives, e.g., convergence, effectiveness, scalability, and robustness.

\begin{table*}[!htbp]
\center
\caption{Representative DRL algorithms for SBEM}\label{table_4}\centering
\begin{tabular}{|m{1.5cm}<{\centering}|m{1.9cm}<{\centering}||m{2.9cm}<{\centering}|m{2.9cm}<{\centering}|m{2.9cm}<{\centering}|m{2.75cm}<{\centering}|}
\hline
\multirow{7}*{\shortstack{~\\~\\~\\~\\~\\~\\~\\~\\~\\~\\~\\~\\~\\~\\~\\~\\~\\~\\~\\~\\~\\~\\~\\ Model-free}} & Description &\multicolumn{4}{c|}{Agents learn optimal policies based on the interaction information with unknown environment} \\
\cline{2-6}
&  Advantages &\multicolumn{4}{c|}{No need to know building environment model, high data-efficiency}\\
\cline{2-6}
&  Disadvantages &\multicolumn{4}{c|}{Require a large number of samples, long exploration time, and high exploration cost}\\
\cline{2-6}
&  \multirow{3}*{\shortstack{~\\~\\Representative ~\\algorithms, ~\\features and ~\\categories}} &DQN &DDPG &PPO &A2C/A3C\\
\cline{3-6}
& &Support only for discrete action space &Support only for continuous action space &Support stable training and discrete/continuous action space &Support fast training and discrete/continuous action space\\
\cline{3-6}
& &Value-based, off-policy &Policy-based, off-policy &Policy-based, on-policy & Policy-based, on-policy\\
\cline{2-6}
& Application scenarios &HVAC control~\cite{Wei2017}\cite{RanYoon2019}\cite{Gupta2020}\cite{Sakuma2020}, MES optimization~\cite{Mocanu2019}\cite{Wei2020}, Microgrid energy management~\cite{Fran2016}\cite{DominguezBarbero2020,ChenTianyi2019,Ji2019} &HVAC control~\cite{Gao2020}, MES optimization~\cite{YuIoT2019}\cite{Ye2020}, Microgrid energy management~\cite{LeiTan2020}\cite{YangXiaoDong2019} &MES optimization~\cite{ZhangChi2019}, Microgrid energy management~\cite{LeeJ2020} &HVAC control~\cite{Morinibu2019}, Microgrid energy management~\cite{Yang2019}\\
\hline
\multirow{7}*{\shortstack{~\\~\\~\\~\\~\\~\\~\\~\\~\\~\\~\\~\\~\\~\\~\\~\\~\\~\\~\\ Model-based}} & Description &\multicolumn{4}{c|}{Agents construct a stimulated environment model and use it to generate future episodes for training} \\
\cline{2-6}
& Advantages &\multicolumn{4}{c|}{High sample efficiency}\\
\cline{2-6}
& Disadvantages &\multicolumn{4}{c|}{Develop an accurate and useful model is often challenging}\\
\cline{2-6}
&  \multirow{3}*{\shortstack{~\\~\\~\\Representative ~\\algorithms, ~\\features and ~\\categories}} &MuZero &LSTM-DDPG &Differentiable MPC-PPO &BEM-A3C\\
\cline{3-6}
& &Learn a network model with accurate planning performance &Use LSTM and historical data to learn transition function and reward function &Pre-train a differentiable MPC policy based on imitation learning, which is both sample-efficient and interpretable &Use EnergyPlus and measured data to simulate building environment \\
\cline{3-6}
& &Model-based, off-policy &Model-based, off-policy &Model-based, on-policy & Model-based, on-policy\\
\cline{2-6}
& Application scenarios &Microgrid energy management~\cite{Shuai2020} &HVAC control~\cite{Zou2019} &HVAC control~\cite{Chen2019} &HVAC control~\cite{Zhang2019} \\
\hline
\end{tabular}
\end{table*}

\subsection{Representative DRL Algorithms for SBEM}
In this subsection, we introduce some representative DRL algorithms for SBEM in Table~\ref{table_4}. According to the descriptions in Section~\ref{s2}-C, these algorithms can be divided into two types, i.e., model-free algorithms and model-based algorithms.

Model-free algorithms do not require explicit building environment models, but they need to collect sufficient experience transitions for training, which may result in a long exploration time and a high exploration cost. In Table~\ref{table_4}, four representative model-free DRL algorithms for SBEM are given, i.e., DQN, DDPG, PPO, and A2C/A3C. To be specific, DQN is a valued-based off-policy algorithm and only supports discrete action space, while DDPG is a policy-based off-policy algorithm and only supports continuous action space. Since experience replay is adopted by DQN and DDPG, they have higher data efficiency compared with ``on-policy" algorithms. However, they tend to overestimate Q-value and generate sub-optimal policies. As two representative policy-based off-policy algorithms, PPO and A2C/A3C can support both discrete and continuous action spaces. Moreover, PPO can support stable learning by controlling the similarity between the current policy and the old policy. Furthermore, it is robust to hyperparameters and network architectures. Although A2C/A3C can support reliable and parallel learning on a single multi-core CPU, it is sensitive to the employed hyperparameters. To show the differences among these algorithms clearly, their network architectures are illustrated in Fig.~\ref{fig_3}.

Model-based algorithms need to construct models to simulate building environments and use them to generate future episodes for training. Therefore, model-based algorithms outperform model-free algorithms in terms of sample complexity. However, for model-based algorithms, it is often challenging to obtain an accurate building environment model. In existing works, many model-based DRL methods for SBEM have been proposed by constructing a system dynamics model with historical operational data\cite{Zou2019}\cite{Chen2019} or calibrating a building environment simulation model with the measured data\cite{Zhang2019}. Four representative model-based DRL algorithms for SBEM are provided in Table~\ref{table_4}, i.e., MuZero, LSTM-DDPG, Differentiable MPC-PPO, and BEM-A3C. To be specific, MuZero is a model-based off-policy algorithm\cite{Latif2021}, which intends to learn a network model with accurate planning performance. Since it uses tree-based search methods, MuZero may not be good at dealing with continuous action space. LSTM-DDPG is also a model-based off-policy algorithm, which uses LSTM and historical operational data to learn the environment model. Then, the obtained model is used to generate a large number of data for training DRL agents with DDPG algorithm. Differentiable MPC-PPO is a model-based on-policy algorithm, which uses differentiable MPC to learn existing controller via imitation learning and improves the learned controller via PPO algorithm. BEM-A3C is also a model-based on-policy algorithm, which uses EnergyPlus software to develop a building energy model and uses actual operational data to calibrate the model. Finally, the calibrated model can be used for training DRL agents via A3C algorithm.

\begin{figure}[!htb]
\centering
\includegraphics[scale=1]{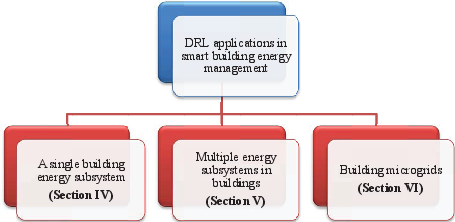}
\caption{Taxonomy of DRL Applications in SBEM}\label{fig_4}
\end{figure}

In the next three sections, we will introduce DRL applications in SBEM considering different system scales as shown in Fig.~\ref{fig_4}, i.e., a single building energy subsystem, multiple energy subsystems (MES) in buildings, and building energy systems in microgrid environment.

\section{Applications of DRL in a Single Building Energy Subsystem}\label{s4}
In existing works, DRL techniques have been adopted to optimize the operation cost or energy consumption of a single building energy subsystem. Among all single building energy subsystems, HVAC and EWH\cite{Kazmi2018,Ruelens2019,Peirelinck2021} have very flexible power consumption\cite{Minoli2017}. Therefore, we mainly focus on them in this section. To be specific, model-free DRL methods for HVAC control, model-based DRL methods for HVAC control, and DRL methods for EWHs are introduced in section~\ref{s41}, section~\ref{s42}, and section~\ref{s43}, respectively. At the end of this section, we give a summary of existing works and provide some insights.

\subsection{Model-free DRL Methods for HVAC Control}\label{s41}
It is well known that the main purpose of an HVAC system is to maintain thermal comfort for occupants. To achieve this aim, many DRL-based methods have been proposed. For example, Morinibu \emph{et al.} proposed an A2C-based method to decrease the non-uniformity of radiation temperature in a room by flexibly controlling the wind direction of an HVAC system\cite{Morinibu2019}. Simulation results showed that the proposed method has better performance than random control and normal control. Since the operations of HVAC systems place an economic burden on building operators, it is very necessary to minimize energy cost while maintaining thermal comfort for occupants. In \cite{Wei2017}, Wei \emph{et al.} proposed a DQN-based HVAC control method to save energy cost in office buildings while maintaining the room temperature requirements. When 5 zones are considered, energy cost can be reduced by 35.1\%. Similarly, Nagy \emph{et al.} and Gupta \emph{et al.} proposed a model-free DRL-based HVAC control method in a residential building to save energy cost and reduce the loss of occupant comfort based on D-DNFQI and DQN, respectively\cite{Nagy2018}\cite{Gupta2020}. In addition to energy cost, energy consumption is also an important metric. To minimize energy consumption while maintaining thermal comfort, some works have been done in \cite{RanYoon2019}\cite{Sakuma2020}. Since the above-mentioned works use value-based DRL methods, they can not deal with continuous actions. To support continuous actions, Gao \emph{et al.} presented a DDPG-based HVAC control method to optimize energy consumption and thermal comfort in a laboratory by jointly adjusting temperature set-point and humidity\cite{Gao2020}. Simulation results showed that the proposed method has higher thermal comfort and energy-efficiency than other baselines, e.g., Q-learning and DQN. Due to the importance of indoor air quality, Valladares \emph{et al.} proposed a DDQN-based control algorithm to optimize HVAC energy consumption while maintaining thermal comfort and indoor air quality comfort for occupants\cite{Valladares2019}. Although the above-mentioned methods are effective, they can not be used for coordinating multiple components in HVAC systems.

\begin{figure}[!htb]
\centering
\includegraphics[scale=0.6]{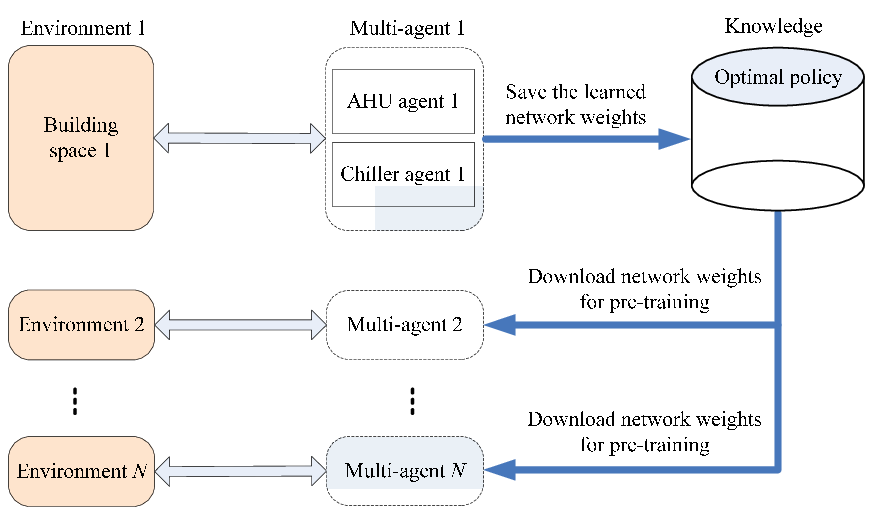}
\caption{The proposed transfer learning framework for multi-agent training}\label{fig_5}
\end{figure}

To overcome this drawback, Nagarathinam \emph{et al.}\cite{Nagarathinam2020} proposed a multi-agent DRL based algorithm to minimize HVAC energy consumption without sacrificing user comfort by adjusting both the building and chiller set-points. To be specific, each DDQN-based agent coordinates with others to learn an optimal HVAC control policy. Note that the coordination is achieved by allocating the same reward for each agent. Since a large building may have a few hundreds of AHUs and a few tens of chillers, it is time-consuming to train all agents centrally. To achieve an accelerated learning process, transfer learning (i.e., transferring the knowledge from one task to a related but different task\cite{Taylor2009}\cite{Zhangx2020}) is adopted. As shown in Fig.~\ref{fig_5}, the optimal policies obtained by training multiple agents on a sub-set of HVAC systems (including one AHU and one chiller) can be used to pre-train multiple agents related to other HVAC subsystems due to the problem similarity.

\subsection{Model-based DRL Methods for HVAC Control}\label{s42}

Although the above-mentioned works are effective, there are two drawbacks in the process of training a DRL agent. Firstly, it is impractical to let the DRL agent to explore the state space fully in a real building environment since unacceptably high cost may be incurred\cite{Mason2019}\cite{Zou2019}\cite{Chen2019}. Secondly, it may take a long time for the DRL agent to learn an optimal policy if trained in a real-world environment\cite{Zou2019}\cite{Chen2019}.

\begin{figure}[!htb]
\centering
\includegraphics[scale=0.57]{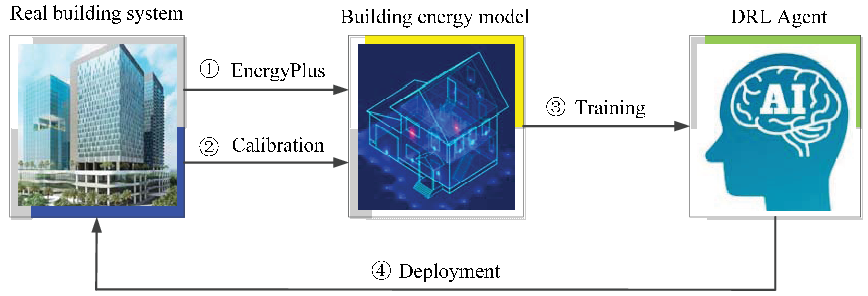}
\caption{BEM-based DRL control framework}\label{fig_6}
\end{figure}

To reduce the number of interactions with a real building environment, many model-based DRL control methods have been developed\cite{Zou2019}\cite{Zhang2019}. For example, Zhang \emph{et al.}\cite{Zhang2019} proposed and implemented a building energy model (BEM)-based DRL control framework for a novel radiant heating system in an existing office building. The proposed framework consists of four steps as shown in Fig.~\ref{fig_6}, i.e., building energy modeling, model calibration, DRL training, and real deployment. To be specific, EnergyPlus software is used to develop a building energy model for the office building. Next, based on the observed data, the building energy model can be calibrated. Then, the calibrated model is used as the simulator of environment for training the DRL agent off-line based on A3C algorithm. Finally, the learned control policy will be deployed in building automation system (BAS) for generating control signals in real-time. Experimental results showed that the obtained control strategy can reduce heating demand by 16.7\% compared with the rule-based control strategy.

\begin{figure}[!htb]
\centering
\includegraphics[scale=0.65]{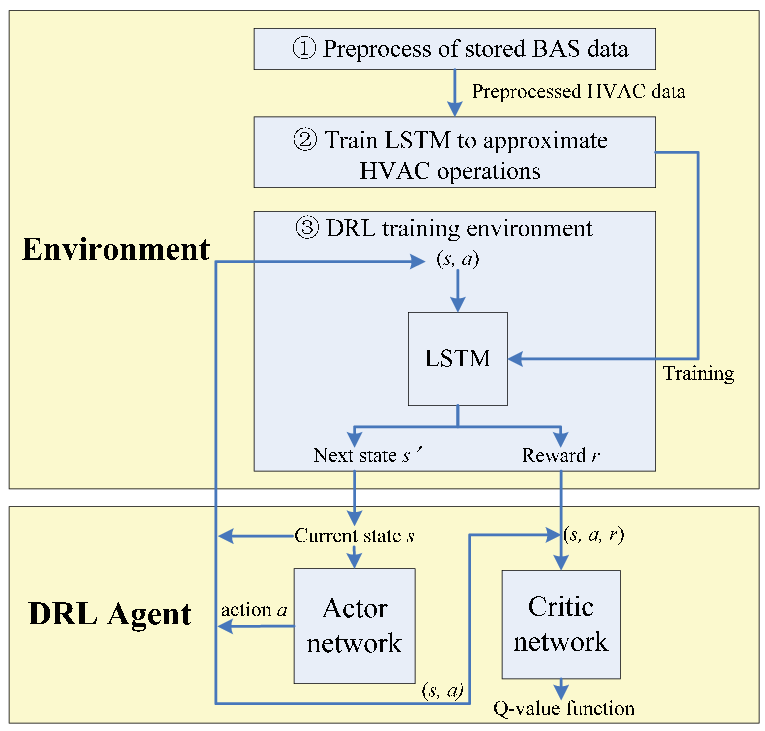}
\caption{LSTM-based DRL control framework}\label{fig_7}
\end{figure}

In \cite{Zhang2019}, the real-world HVAC operational data in three months are used for calibrating building energy models, which will affect model accuracy. To overcome this drawback, Zou \emph{et al.}\cite{Zou2019} proposed a DRL-based HVAC control framework to minimize energy consumption while maintaining thermal comfort levels for occupants based on operational data within two years. The proposed framework is composed of two parts as shown in Fig.~\ref{fig_7}, i.e, creating DRL training environment and training DRL agent based on the created environment. To be specific,
LSTM models are built based on BAS historical data, which can approximate HVAC operations. Note that the inputs of LSTM models are current state and action, while their outputs are next state and reward. After LSTM networks are trained, they can be used to create training environment. Next, DRL agent interacts with the training environment until it converges to an optimal HVAC control policy. Finally, the optimal control policy can be deployed for controlling AHUs in real-time. Moreover, DRL agent contains an actor network and a critic network, which are trained using DDPG algorithm. Algorithmic testing results showed that DRL agents can save energy by 27\% to 30\% while maintaining the predicted percentage of discomfort at 10\%.

\begin{table*}[htbp]
\center
\caption{Summary of Existing Works on DRL for a Single Building Energy Subsystem}\label{table_5} \centering
\begin{tabular}{|m{2.85cm}<{\centering}|m{1.6cm}<{\centering}|m{1.2cm}<{\centering}|m{1.8cm}<{\centering}|m{2cm}<{\centering}|m{1.7cm}<{\centering}|m{1.8cm}<{\centering}|m{1.6cm}<{\centering}|}
\hline
\textbf{Research work}& \textbf{Object(s)} & \textbf{Subsystem} & \textbf{Primary objective}& \textbf{Secondary objective(s)}& \textbf{DRL algorithm, function estimator} & \textbf{Performance improvement} &\textbf{Practical implementation}\\
\hline
\hline
Morinibu \emph{et al.}\cite{Morinibu2019}&	Smart home&	HVAC & Non-uniformity of radiant temperature& Thermal comfort& A2C, CNN\&LSTM & ---  &No\\
\hline
Wei \emph{et al.}\cite{Wei2017}& Office&	HVAC&  Energy cost&	Thermal comfort& DQN, DNN &19.1\%$\sim$ 71.2\%  &No\\
\hline
Nagy \emph{et al.}\cite{Nagy2018}& Residential buildings&	HVAC&  Energy cost& Thermal comfort& D-DNFQI, DNN & 5.5\%$\sim$10\%  &No\\
\hline
Gupta \emph{et al.}\cite{Gupta2020}&	Residential buildings&	HVAC& Energy cost& Thermal comfort & DQN, DNN &5\%$\sim$12\% & No\\
\hline
Yoon \emph{et al.}\cite{RanYoon2019}&	Office&	HVAC&	 Energy consumption& Thermal comfort& DQN, DNN &12.4\%$\sim$32.2\% &No\\
\hline
Sakuma \emph{et al.}\cite{Sakuma2020}&	Residential buildings&	HVAC& Energy consumption& Thermal comfort & DQN, DNN &34.5\% & No\\
\hline
Gao \emph{et al.}\cite{Gao2020} &	Laboratory&	HVAC&		Energy cost	& Thermal comfort& DDPG, DNN &4.31\%$\sim$9.15\% & No\\
\hline
Valladares \emph{et al.}\cite{Valladares2019}&	Laboratory and classroom&	HVAC& Energy cost& Thermal comfort, air quality& DDQN, DNN &4\%$\sim$5\% & No\\
\hline
Nagarathinam \emph{et al.}\cite{Nagarathinam2020} & A campus building&	HVAC & Energy consumption & Thermal comfort & Multi-agent DDQN, DNN & 17\% & No \\
\hline
Zhang \emph{et al.}\cite{Zhang2019}& Office&	HVAC&	Energy cost& Thermal comfort& BEM-A3C, DNN & 7.06\%$\sim$16.7\% &Yes\\
\hline
Zou \emph{et al.}\cite{Zou2019}& Office&	HVAC& 	Energy consumption& Thermal comfort& LSTM-DDPG, LSTM  & 27\%$\sim$31.27\%& No\\
\hline
Chen \emph{et al.}\cite{Chen2019}& Office&	HVAC&		Energy cost& Thermal comfort& Differentiable
MPC-PPO, Linear model & 16.7\% & Yes\\
\hline
Kazmi \emph{et al.}\cite{Kazmi2018}& Residential buildings&	EWH &		Energy consumption& Thermal comfort, exploration bonus& Deep PILCO, DNN & 20\% & Yes\\
\hline
Ruelens \emph{et al.}\cite{Ruelens2019}& Residential buildings&	EWH &		Energy cost& Thermal comfort& Fitted Q-iteration, CNN/LSTM & 5.5\%$\sim$10.2\% & No\\
\hline
Peirelinck \emph{et al.}\cite{Peirelinck2021}& Residential buildings&	EWH&		Energy cost& Thermal comfort& DDQN, DNN &8.8\%$\sim$32.2\% & No\\
\hline
\end{tabular}
\end{table*}

\begin{figure}[!htb]
\centering
\includegraphics[scale=0.64]{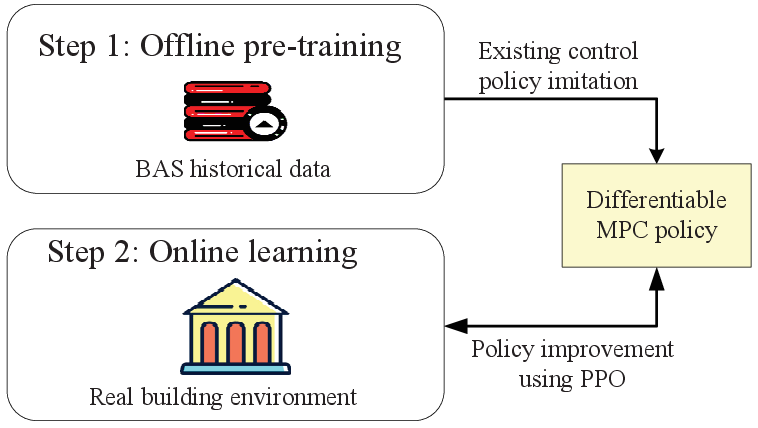}
\caption{Differentiable MPC policy based HVAC control framework}\label{fig_8}
\end{figure}

Similar to \cite{Zou2019}, Chen \emph{et al.} proposed a PPO-based approach for HVAC control by utilizing historical data so that practical deployment can be achieved\cite{Chen2019}. To be specific, the framework of the proposed approach is shown in Fig.~\ref{fig_8}. Firstly, historical data from existing HVAC controllers are used to pre-train a differentiable MPC policy based on imitation learning. Note that the pre-trained policy can encode domain knowledge into planning and system dynamics, making it both sample-efficient and interpretable. Secondly, the pre-trained control policy is improved continually in the process of interacting with the real building environment using online learning algorithm. Since PPO is robust to hyperparameters and network architectures, it is adopted to improve the pre-trained policy. Practical experimental results showed that the proposed approach can save 16.7\% of cooling demand compared with the existing controller and track temperature set-point better.

\subsection{DRL Methods for EWH Control}\label{s43}
In an EWH, there are many separate layers of water and each layer has a unique temperature in practice, measuring the temperature within the EWH using a single sensor will lead to sparse observations. In \cite{Ruelens2019}, Ruelens \emph{et al.} proposed an effective method to tackle sparsely observed control problem related to EWHs based on fitted Q-iteration and LSTM. The key idea of the proposed method is to store sequences of past observations and actions in the state vector so that relevant features for finding near-optimal control policies can be extracted based on RL. Simulation results showed that LSTM has better performance than CNN and DNN when they are used as function estimators in RL.

Since training a DRL agent without knowing system dynamics of EWHs requires a large number of interactions with the actual environment, model-based DRL methods are preferred due to their high sample efficiency. For example, Kazmi \emph{et al.} proposed a model-based DRL method to optimize the hot water production based on Deep PILCO, which can reduce energy consumption by about 20\% and has been applied to a set of 32 houses in the Netherlands. The key idea of the proposed method is summarized as follows. Firstly, executing actions under the current policy and collecting experience transition data for training system dynamics model. Next, generating trajectories based on the current policy and the obtained dynamics model. Then, trajectories are used to update policy. Finally, the updated policy will be used in next loop.

\begin{figure}[!htb]
\centering
\includegraphics[scale=0.7]{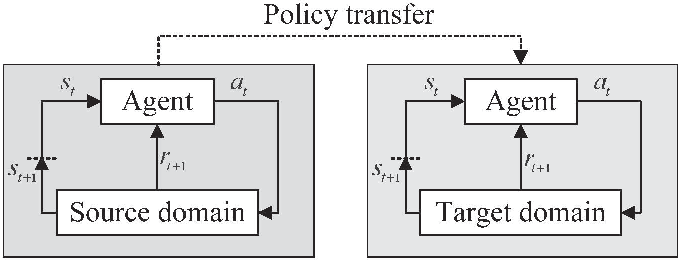}
\caption{Domain randomization based pre-training method}\label{fig_9}
\end{figure}

When prior knowledge about system dynamics is available, learning a model from observations could be avoided. In \cite{Peirelinck2021}, Peirelinck \emph{et al.} investigated an energy cost minimization problem related to EWHs given a known system dynamics model. Since the model expression is very complex, it is challenging to find the optimal policy for EWH operation. Therefore, DRL is used to obtain the optimal policy since it merely cares about the input and output of the model. To reduce the training time in target domain (i.e., practical environment), domain randomization is used as shown in Fig.~\ref{fig_9}. To be specific, model parameters are randomized in the source domain (i.e., training environment). Based on the model with uncertain parameters, a generalized policy in the source domain is trained. Then, the trained policy is transferred to target domain for initializing DNNs. Simulation results showed that pre-training is helpful for reducing energy cost by 8.8\%.

\textbf{Summary}: In this section, we review existing works on DRL for a single building energy subsystem. For easy reading, the specific details including objectives, DRL algorithms, and implementation methods are summarized in Table~\ref{table_5}. It can be observed that most existing works focus on HVAC control since HVAC systems have the largest energy consumption among all single building energy subsystems\cite{Minoli2017}. Moreover, most of optimization objectives are related to energy cost/consumption and thermal comfort. By controlling an HVAC system intelligently based on DRL methods, its energy cost can be reduced by 4\%-71.2\% and energy consumption can be decreased by 12.4\%-34.5\% without sacrificing thermal comfort. In addition, nearly all model-free DRL methods are evaluated by simulations and several model-based DRL methods have been deployed in practice.

\section{Applications of DRL in Multiple Energy Subsystems of Buildings}\label{s5}
In this section, we will introduce the applications of DRL in multiple energy subsystems of residential buildings and commercial buildings, respectively. To be specific, section~\ref{s51} focuses on the coordination of home energy management system, HVAC systems, ESSs, EVs, WMs, PVs, and EWHs in residential buildings, while section~\ref{s52} focuses on the coordination of HVAC systems, lighting systems, blind systems, window systems, and personal electric devices in commercial buildings. Moreover, we give a summary of existing works and provide some insights in the last paragraph of this section.

\subsection{Multiple Energy Subsystems in Residential Buildings}\label{s51}
As the smallest unit in a residential building, smart home has many kinds of appliances, e.g., HVAC systems, EVs, ESSs, and PVs. To implement the coordination of different appliances, many DRL-based methods have been proposed to save energy cost. For example, Yu \emph{et al.} proposed a DDPG-based home energy management algorithm to minimize energy cost for the joint scheduling of HVAC systems and ESSs\cite{YuIoT2019}. Simulation results showed that the proposed algorithm can reduce energy cost by 8.1\%-15.21\% through the utilization of temporal diversity of dynamic prices\cite{Wan2018}\cite{Kumar2019}. Similar works can be found in \cite{Liu2019} and \cite{He2020}. To be specific, DDQN and TRPO based methods have been proposed to minimize energy cost of a smart home with the consideration of occupant satisfaction degree or thermal comfort, respectively.

As for residential buildings, other objectives may be pursued when optimizing its energy use, e.g., peak demand\cite{Mocanu2019}, transformer capacity violation\cite{ZhangChi2019}, and revenue of excess renewable energy\cite{Ye2020}. For example, Mocanu \emph{et al.} proposed two algorithms to minimize energy cost and peak load of residential buildings with the consideration of HVAC systems, EVs, and dishwashers (DWs)\cite{Mocanu2019}. Simulation results illustrated that the proposed algorithms based on DQN and deep policy gradient can efficiently cope with the inherent uncertainty and variability in renewable energy generation and power demand. Although the proposed algorithms in \cite{Mocanu2019} are effective, they neglect the physical constraints related to residential buildings, e.g., transformer capacity. To deal with this limitation, Zhang \emph{et al.} investigated a multi-household energy management problem for residential units connected to the same transformer. Since violating the transformer capacity is harmful to its lifetime, an efficient approach was designed based on cooperative multi-agent DRL\cite{ZhangChi2019}. Simulation results indicated that the energy cost of residential households can be reduced by 59.77\% without violating the transformer capacity.

\begin{figure}[!htb]
\centering
\includegraphics[scale=0.54]{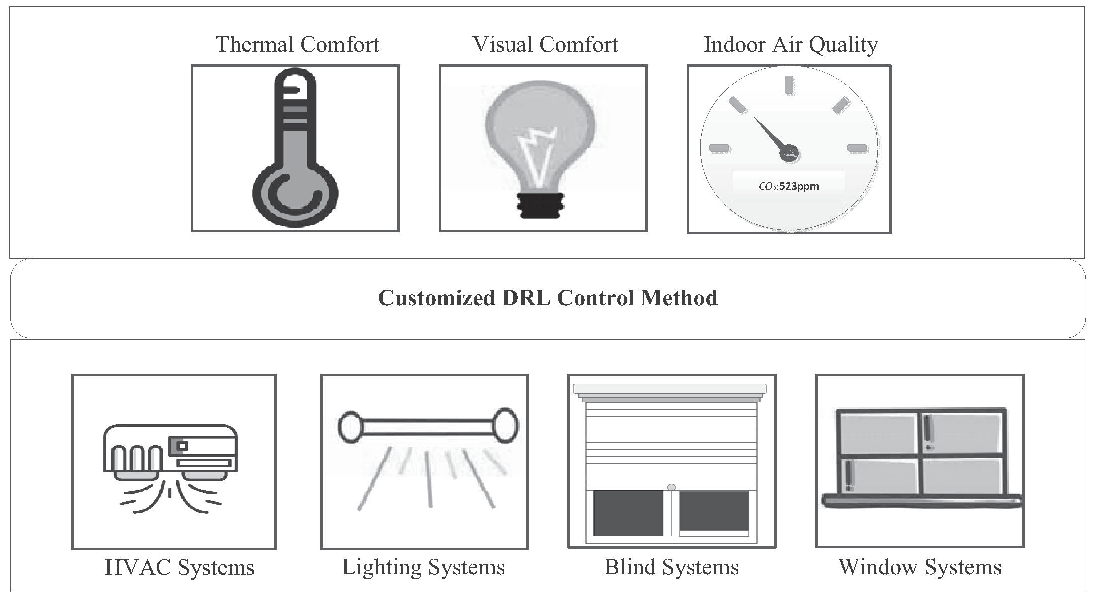}
\caption{The architecture of the proposed control framework}\label{fig_10}
\end{figure}

\subsection{Multiple Energy Subsystems in Commercial Buildings}\label{s52}
In existing works, some DRL-based approaches have been proposed to reduce energy consumption in commercial buildings \cite{Zou2019}\cite{Zhang2019}\cite{Valladares2019}. Although some advances have been made, these works only consider a single subsystem in
buildings (e.g., an HVAC system) without noticing that other subsystems can also affect energy consumption and user comfort in terms of
thermal, air quality, and illumination conditions. In fact, some research results showed that jointly controlling HVAC systems and other building energy subsystems (e.g., blind systems, lighting systems, and window systems) has great potential of saving energy\cite{Cheng2016}\cite{Wang2015}. For example, HVAC energy consumption can be reduced by 17\%-47\% if window-based natural ventilation is adopted\cite{Wang2015}. Based on the above observation, Ding \emph{et al.} proposed a DRL-based framework as shown in Fig.~\ref{fig_10} for efficiently controlling four building energy subsystems (including HVAC systems, lighting systems, blind systems, and window systems\cite{Ding2019}) so that the total energy consumed by all subsystems can be minimized while still maintaining user comfort. To solve the high-dimensional action problem, a branching dueling Q-Network (BDQ) algorithm was used. Moreover, a calibrated EnergyPlus simulation model was adopted to generate enough data for the training of the DRL agent. Simulation results showed that the proposed framework can save energy by 14.26\% compared with the rule-based method while maintaining human comfort within a desired range.

\begin{figure}[!htb]
\centering
\includegraphics[scale=0.45]{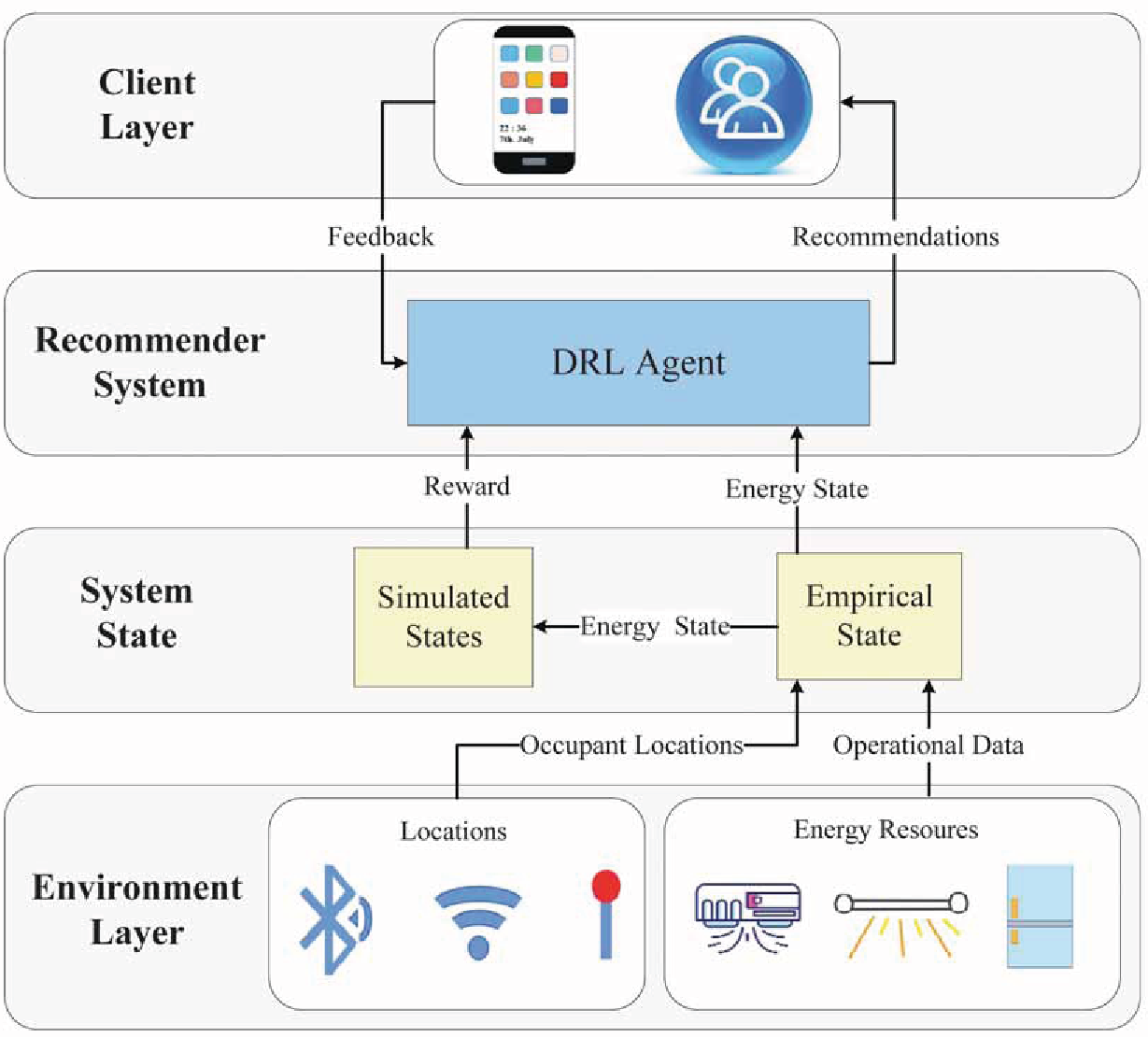}
\caption{The system architecture of the designed recommender}\label{fig_11}
\end{figure}

However, the above-mentioned works mainly focus on building energy system itself and treat occupants as immovable objects, which may decrease the potential of reducing energy consumption and be illustrated by the following example. Suppose that a space is sparsely occupied by some occupants. If they are recommended to vacate this space and move to another occupied space, their comforts may be not sacrificed while the energy consumption in the vacated space can be reduced. Therefore, it is very necessary to investigate the potential of saving energy by shaping occupant behavior. To this end, Wei \emph{et al.}\cite{Wei2020} designed a DRL-based recommender system in commercial buildings, which can learn actions with high energy saving potential and distribute recommendations to occupants. Based on the feedback from occupants, better recommendations can be learned. The system architecture of the designed recommender is shown in Fig.~\ref{fig_11}, which consists of four layers, i.e., \emph{environment layer}, \emph{system state layer}, \emph{recommender system}, and \emph{client layer}. To be specific, \emph{environment layer} measures building environment (e.g., occupant locations and energy consumption information) and sends such information to \emph{system state layer}. \emph{System state layer} contains two components, i.e., an empirical state, which maintains the current building state, and simulated states, which are used to represent the next state after the potential energy saving actions are taken. \emph{The recommender system layer} learns the potential of different recommendation actions (including \emph{move recommendation}, \emph{schedule change}, \emph{reduce personal resources}, and \emph{reduce service in spaces}). \emph{The client layer} receives recommendations and allows clients to provide feedback (e.g., accept or reject the recommendation). A four-week user study showed that the designed recommender system can reduce building energy consumption by 19\% to 26\% compared with a passive-only strategy.

\begin{table*}[htbp]
\center
\caption{Summary of Existing Works on DRL for Multi-energy Subsystems in Buildings}\label{table_6} \centering
\begin{tabular}{|m{2.15cm}<{\centering}|m{1.65cm}<{\centering}|m{1.9cm}<{\centering}|m{1.7cm}<{\centering}|m{2cm}<{\centering}|m{1.5cm}<{\centering}|m{1.7cm}<{\centering}|m{1.7cm}<{\centering}|}
\hline
\textbf{Research work}& \textbf{Object(s)} & \textbf{Energy subsystems}& \textbf{Primary objective}& \textbf{Secondary objective(s)}& \textbf{DRL algorithm, function estimator} & \textbf{Performance improvement} & \textbf{Practical implementation}\\
\hline
\hline
Yu \emph{et al.}\cite{YuIoT2019}&	Smart home&	PV, ESS, HVAC&	Energy cost& Thermal comfort& DDPG, DNN & 8.10\%$\sim$15.21\% & No\\
\hline
Liu \emph{et al.}\cite{Liu2019} &	Smart home&	PV, ESS, HVAC, EV, Heater, DW &	Energy cost	& Consumers' satisfaction degree& DDQN, DNN & 41.8\%$\sim$59\% & No\\
\hline
Li \emph{et al.}\cite{He2020}&	Smart home&	HVAC, EV, EWH, DW, WM&	Energy cost& Thermal comfort and range anxiety&	 TRPO, DNN & 31.6\% & No\\
\hline
Mocanu \emph{et al.}\cite{Mocanu2019}& Residential buildings& PV, HVAC, EV, DW & Energy cost &	Peak demand, load operational time or condition&	DQN, DNN & 14.1\%$\sim$27.4\% & No\\
\hline
Zhang \emph{et al.}\cite{ZhangChi2019}& Residential buildings& PV, ESS, EV& Energy cost& Transformer capacity violation & PPO, DNN & 59.77\% & No\\
\hline
Ye \emph{et al.}\cite{Ye2020}& Residential buildings&	PV, ESS, TES, EHP, GB&	Energy cost& Excess energy sale revenue& PDDPG, DNN & 6.28\%$\sim$10.21\%& No\\
\hline
Ding \emph{et al.}\cite{Ding2019} &Commercial buildings & HVAC, lighting, blind and window & Energy consumption & Thermal comfort, IAQ, lighting comfort & BDQ, DNN & 14.26\% & No\\
\hline
Wei \emph{et al.}\cite{Wei2020} & Commercial buildings & HVAC, lighting, plug load & Energy consumption & Safety, comfort, productivity & DQN, DNN &19\%$\sim$26\% & Yes \\
\hline
\end{tabular}
\end{table*}

\textbf{Summary}: In this section, we review existing works on DRL applications in multiple energy subsystems of buildings. For easy understanding, the research objects, considered energy subsystems, research objectives, DRL algorithms, performance improvement, and implementation methods in existing works are summarized in Table~\ref{table_6}. It can be observed that there is a great potential in reducing energy cost of buildings by scheduling multiple energy subsystems coordinately, e.g., relative energy cost reduction is up to 59\% while maintaining comfort of occupants. Compared with the optimal HVAC control in Table~\ref{table_5}, more advanced DRL algorithms are adopted to deal with more complex problems, e.g., PDDPG, BDQ, and TRPO. In addition, most of DRL methods are evaluated by simulations.

\section{Applications of DRL in Building microgrids}\label{s6}
In this section, we review the existing works on DRL-based energy optimization for building microgrids. To be specific, section~\ref{s61} introduces DRL-based energy management algorithms for microgrids with uncontrollable building loads, while section~\ref{s62} introduces DRL-based microgrid optimization algorithms considering the flexibility of building loads. Moreover, we summarize the existing works and point out some insights at the end of this section.

\subsection{Microgrid Optimization without Considering Controllable Building Loads}\label{s61}
In existing works, many DRL-based methods have been proposed for residential microgrids\cite{Fran2016,DominguezBarbero2020,ChenTianyi2019,Ji2019,Shuai2020}, where a microgrid is a low voltage distribution network comprising various distributed generation, storage devices, and responsive loads\cite{Liu2017}. For example, Francois-Lavet \emph{et al.} proposed a DQN-based control algorithm for a residential microgrid with the consideration of battery and hydrogen storage device to minimize the levelized energy cost\cite{Fran2016}. Similar work can be found in \cite{Ji2019}. However, power demand is assumed to be satisfied in \cite{Fran2016}. For an isolated residential microgrid, load shedding may happen when the total power supply is smaller than the total power demand. At this time, non-served power demand should be penalized. Based on this observation, Dominguez-Barbero \emph{et al.} proposed a DQN-based microgrid optimization algorithm to minimize the sum of DG generation cost and the penalty of non-served power demand\cite{DominguezBarbero2020}. Different from the above works, Chen \emph{et al.} investigated a peer-to-peer energy trading problem among multiple microgrids\cite{ChenTianyi2019}. Moreover, a DQN-based energy trading strategy was proposed to maximize the utility function in a microgrid, which is related to trading profit, retail profit, battery wear cost, demand penalty, and virtual penalty. Simulation results based on one-year real generation and demand data showed the effectiveness of the proposed strategy. Although some advances have been made in above efforts, the proposed DQN-based methods can not deal with DRL problems with continuous actions (e.g., the generation output of DGs\cite{LeiTan2020}).

\begin{figure}[!htb]
\centering
\includegraphics[scale=0.45]{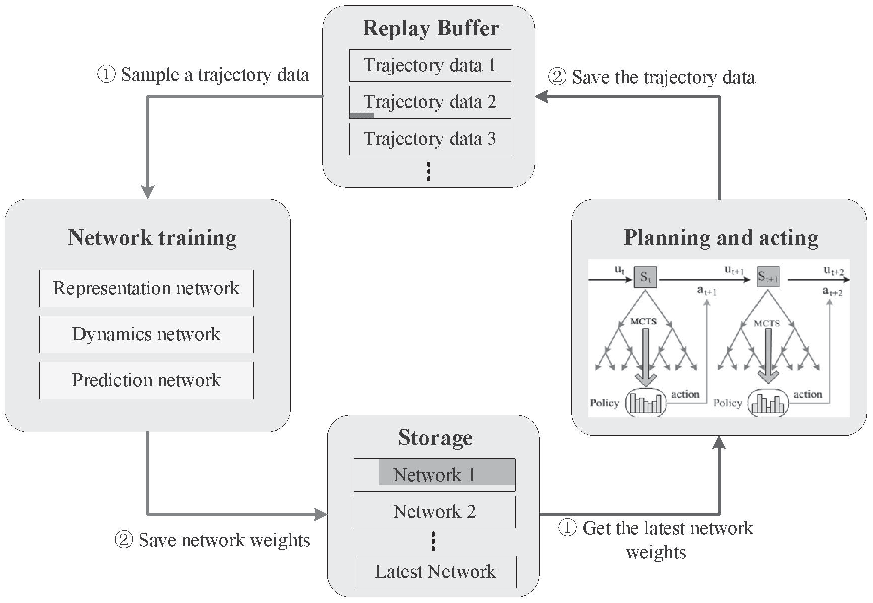}
\caption{The training process of the network model}\label{fig_12}
\end{figure}

To support continuous actions, DDPG-based methods could be adopted. For example, Lei \emph{et al.} proposed a FH-DDPG based energy management algorithm for an isolated microgrid to minimize the sum of power generation cost and the power unbalance penalty\cite{LeiTan2020}. Since model-free based DRL algorithms in existing works have low sample efficiency, Shuai \emph{et al.} proposed a model-based DRL algorithm (i.e., \emph{MuZero}) for the online scheduling of a residential microgrid under uncertainties\cite{Shuai2020} based on Monte-Carlo tree search (MCTS) strategy with a learned network model. Note that the off-line learning process of the network model can be depicted by Fig.~\ref{fig_12}, where four components can be identified, i.e., \emph{network training}, \emph{replay buffer}, \emph{storage}, \emph{planning and acting}. Firstly, the latest network weights are obtained from a storage and used for planning implemented by MCTS. Next, an action is sampled from the search policy, which is proportional to the visit count for each action from the root node. Then, the environment returns a new state and a reward. At the end of the episode, the trajectory data is stored into a replay buffer. When conducting network training, a trajectory data will be randomly sampled from the replay buffer and the updated network weights will be saved in a storage device. It is obvious that network training and trajectory data generation are two independent processes, which can be implemented in parallel. Once the training process of the network model (including three components, i.e., \emph{representation}, \emph{dynamics}, and \emph{prediction}) is completed, the learned network model can be used as the simulator of the real environment. Based on MCTS and the network model, the optimal policy can be learned. Note that the proposed algorithm can operate without relying on any forecasting information and statistic distribution information of the system.

\begin{table*}[htbp]
\center
\caption{Summary of Existing Works on DRL for Microgrids}\label{table_7} \centering
\begin{tabular}{|m{2.35cm}<{\centering}|m{1.6cm}<{\centering}|m{2cm}<{\centering}|m{1.5cm}<{\centering}|m{2cm}<{\centering}|m{1.5cm}<{\centering}|m{1.6cm}<{\centering}|m{1.6cm}<{\centering}|}
\hline
\textbf{Research work}& \textbf{Microgrid type} &  \textbf{Energy systems} &\textbf{Controllable building load considered} & \textbf{Optimization objective(s)}&\textbf{DRL algorithm, function estimator}&\textbf{Cost reduction}&\textbf{Practical implementation}\\
\hline
\hline
Francois-Lavet \emph{et al.}\cite{Fran2016}& A residential microgrid  & PV, Battery, hydrogen storage device & No & The overall levelized energy cost& DQN, DNN &5\%$\sim$12\% & No\\
\hline
Ji \emph{et al.}\cite{Ji2019}& A residential microgrid   & PV, WT, DG, ESS  & No & Daily operating cost & DQN, DNN & 20.75\% & No\\
\hline
Dominguez-Barbero \emph{et al.}\cite{DominguezBarbero2020}& An isolated residential microgrid & PV, DG, Battery, hydrogen storage device & No & Operating cost & DQN, DNN &58.5\%$\sim$67.20\% & No\\
\hline
Chen \emph{et al.}\cite{ChenTianyi2019}& A residential microgrid  &  PV, ESS  & No & Profit minus cost & DQN, DNN & $>$30\% & No\\
\hline
Lei \emph{et al.}\cite{LeiTan2020}& An isolated microgrid   & PV, DG, ESS & No & Power generation cost, power unbalance & FH-DDPG/RDPG, LSTM & 80\% & No\\
\hline
Shuai \emph{et al.}\cite{Shuai2020}&  A residential microgrid   & PV, WT, ESS  & No & Operating cost & MuZero, DNN\&LSTM & 9.28\%$\sim$28.93\% & No\\
\hline
Yang \emph{et al.}\cite{YangXiaoDong2019}& A data center microgrid   & PV, ESS, servers  & Yes, servers & Energy cost & DDPG, DNN & 6.24\% & No\\
\hline
Yang \emph{et al.}\cite{Yang2019}& A residential microgrid   &  PV, ESS, EV & Yes, EV & Energy cost and peak load & Multi-agent EB-C-A2C/DQN, DNN & 24.69\% & No\\
\hline
Lee \emph{et al.}\cite{LeeJ2020}& A residential microgrid   &  WM, CD, WH, DW and refrigerator. & Yes, household appliances & Energy cost and peak load & Multi-agent PPO, DNN & --- & No\\
\hline
\end{tabular}
\end{table*}

\subsection{Microgrid Optimization with Controllable Building Loads}\label{s62}
In above-mentioned works, building loads are regarded as uncontrollable resources in microgrids. In fact, the energy cost of a microgrid could be reduced by scheduling loads flexibly. For example, Yang \emph{et al.} proposed a DDPG-based scheduling algorithm for a data center microgrid with renewable sources to reduce energy cost by choosing the execution time and the quantity of served workloads flexibly\cite{YangXiaoDong2019}. Simulation results showed that energy cost can be reduced by 6.42\%. However, the microgrid optimization problem would be intractable if the number of controllable resources is large due to the increased action space. At this time, multi-agent DRL may be a good choice, which can coordinate all agents effectively. For example, Yang \emph{et al.} proposed an entropy-based collective multi-agent DRL algorithm to schedule EVs and ESSs in large-scale households. Simulation results based on real-world traces showed the effectiveness of the proposed algorithm in reducing the operating cost and the peak load\cite{Yang2019}. Similarly, Lee \emph{et al.} proposed an MAPPO-based algorithm to solve the demand response problem in a microgrid of residential district\cite{LeeJ2020}. The proposed algorithm intends to train multiple household agents centrally. Once an optimal policy is learned by each household agent, it can schedule household appliances without knowing specific information about other households.

\textbf{Summary}: In this section, we review existing works on DRL applications in building microgrids and summarize the details of existing works in Table~\ref{table_7}. It can be observed that existing works mainly focus on economic impacts of building microgrids and the proposed DRL-based methods can indeed bring economic benefits for microgrid operators. However, most of them neglect the control of building loads and all of them are not implemented in practice.

\section{Open Issues and Future Research Directions}\label{s7}
Although recent years have witnessed the rapid development of DRL for SBEM, there are still some unsolved issues that need better solutions. In this section, we highlight open issues and point out future research directions.

\subsection{Data-efficient building energy optimization}
As mentioned in Sections~\ref{s4}-\ref{s6}, most model-free DRL methods for SBEM are still not be implemented in practice. The main reason for this phenomenon is that DRL agents have to interact with the building environment directly so as to collect enough data for training, which is a time-consuming process. Moreover, in the process of interaction, actions are taken by trial and error, resulting in a high exploration cost. For example, random selection of an HVAC temperature set-point may lead to thermal discomfort and high energy consumption. When mitigating these issues, there are several opportunities. To be specific, with the development of IoT technologies, many sensing equipments can be deployed to collect building operational data. Then, the collected data could be used to train DRL agents in an offline way. Moreover, deep meta reinforcement learning\cite{Huang2021} could be used to implement fast learning using only a few data and training episodes. In addition, the collected operational data can be used to learn an environment model and consequently model-based DRL methods can be used to reduce exploration cost.

\subsection{Multi-timescale building energy optimization}
Most existing DRL-based methods focus on single-timescale building energy optimization problems. In fact, there are many multi-timescale decision problems in the field of building energy optimization. For example, supply air temperature and the ratio of re-use air in a commercial building HVAC system can be adjusted once every hour since the frequent adjustment can cause damage to HVAC components\cite{Aswani2012}. In contrast, supply air rate in each zone can be changed every 10-15 minutes\cite{Kalaimani2016}. When confronted with multi-timescale decision problems, existing DRL-based methods are not applicable. A possible way is to design energy optimization algorithms based on the framework of hierarchical DRL\cite{Kulkarni2016}, which can support multi-timescale DRL problems with delayed rewards. In hierarchical DRL, actions can be divided into two types with different timescales. To be specific, actions with long timescale are first taken in the upper level based on system state. Then, actions with short timescale are taken in the lower level based on system state and the chosen actions in the upper level. By coordinating the actions of upper level and lower level, hierarchical DRL-based methods can explore the environments efficiently.

\subsection{Multi-objective building energy optimization}
As shown in Section \ref{s3}, multiple objectives are pursued by SBEM, e.g., energy cost/consumption minimization, carbon emission minimization, and comfort maximization. Moreover, such objectives are often conflicting with each other. A typical way of dealing with conflicting objectives in existing DRL-based methods is to design a synthetic reward function as a weighted sum of different objectives. Since the weight parameters related to different objectives typically have different units and/or scales, it is very challenging to decide their proper values beforehand. Moreover, the learned policies based on the above-mentioned way can not support flexible operation of building energy systems, e.g., switching flexibly between low-energy-cost mode and high-comfort mode. To avoid deciding weighted parameters for multiple objectives and support flexible operations, a possible way is to design building energy optimization algorithms based on some advanced DRL frameworks (e.g., multi-objective DRL\cite{LiK2020}, and multi-objective meta-DRL\cite{Chen2018}).

\subsection{Multi-zone building energy optimization}
In existing works on building HVAC systems, the proposed DRL-based control methods mainly focus on a single-zone building. In \cite{Wei2017}, Wei \emph{et al.} proposed a heuristic algorithm for variable air volume (VAV) HVAC control in a multi-zone office building and DRL agent for each zone was trained separately. Although the proposed algorithm was effective when 5 zones were considered, it was not scalable due to the lack of multi-zone coordination. In \cite{Hu2020}, Hu \emph{et al.} proposed a MADDPG-based method to decide temperature and humidity setpoints in a four-zone building. Since the input of each critic in MADDPG is the concatenation of state and action information from all agents, the scalability of the MADDPG-based method was not very high. In \cite{YuTSG2021}, Yu \emph{et al.} proposed an MAAC-based VAV HVAC control method for a multi-zone commercial building with the consideration of thermal comfort, indoor air quality comfort, and random occupancy, which can operate effectively when 30 zones were considered. Although the above methods are effective when the number of zones is not large, more scalable multi-agent DRL algorithms are expected since the number of zones in a practical commercial building may exceed one hundred or even larger.

\subsection{Efficient training of DRL agents in multi-building energy optimization}
As introduced in Section~\ref{s3}, model-based DRL methods for building energy optimization are sample-efficient. However, a large amount of historical data should be required when learning building thermal dynamics models. For some buildings, especially brand-new buildings, historical data are very limited. At this time, how to speed up the training of building DRL agents is a very challenging task. To improve this situation, a possible way is to combine DRL with transfer learning\cite{Zhu2020TL}. In the field of building energy management, the transferred knowledge may be building thermal dynamics models\cite{Jiangz2019} or control strategies\cite{XuS2020}. Although some efforts have been made in existing works, they mainly focus on transfer learning problems with simple scenarios, where a small similarity gap exists between source MDP and target MDP related to DRL-based SBEM. When the similarity gap is large (e.g., the dimensions of state spaces and action spaces in two MDPs are different), how to design efficient inter-task mapping function and select proper form of the transferred knowledge is very challenging, especially for multi-agent DRL-based SBEM problems.

\subsection{DRL-based energy optimization for building microgrids}
Due to the high thermal inertia, buildings can be regarded as thermal energy storage units. By incorporating building thermal dynamics into microgrid scheduling\cite{Liu2017} or planning\cite{ZhangSPEC2019}, the operation cost or total annualized cost can be reduced. However, explicit building thermal dynamics models are required in the above works. Although DRL-based methods can operate without knowing them, several challenges have to be addressed. Firstly, both discrete and continuous decision variables (e.g., discrete variables are used for describing the operational states of WMs, HVAC loads, and distributed generators, while continuous variables are used for describing the EV charging/discharging power) exist in the optimal operation problem related to building microgrids, which means that discrete-continuous hybrid actions should be supported by the designed DRL-based algorithms. Secondly, multi-agent DRL energy management algorithms with complex reward components should be designed to efficiently promote the coordination among the microgrid controller and all building energy management systems, since each building has their respective objectives (e.g., comfort requirements) and also needs to participate in optimizing the objective of the microgrid.

\textbf{Remarks}: The above-mentioned DRL techniques for SBEM can be supported by existing IEC energy management standards (e.g., ISO/IEC 15067-3-3-2019). For example, ISO/IEC 15067-3-3-2019 defines some energy management agents and provides their operational modes, e.g., single-agent mode, mesh mode, hierarchical mode, and mixed hierarchical and mesh mode. Correspondingly, algorithms based on single-agent DRL, multi-agent DRL, hierarchical DRL, and multi-agent hierarchical DRL can be adopted and implemented by energy management agents.

\section{Conclusions and Lessons Learned}\label{s8}
In this paper, we reviewed the DRL applications in SBEM with the consideration of different system scales comprehensively. In particular, we summarized the features of different DRL methods for SBEM. Moreover, we provided some insights, identified some unsolved issues, and pointed out potential directions for future research. A few major lessons that we learned from this review are summarized as follows. Firstly, nearly all model-free DRL-based building energy optimization methods are still not implemented in practice due to a long exploration time and a high exploration cost. Secondly, model-based DRL approaches for building energy optimization are more practical than model-free DRL approaches since the former can generate enough training data for DRL agents and reduce the number of interactions with the real environment. When the amount of historical data is not enough in the current environment, transfer learning can be used to pre-train a building thermal dynamics model or policy based on the large amount of historical data in a related, but different building environment. Thirdly, compared with some traditional methods, DRL-based energy management methods have the potential of improving some building performance metrics (e.g., energy cost, peak load, and occupant dissatisfaction degree) simultaneously. Finally, although some advances have been made in existing works, there are still many challenges caused by low data efficiency, multiple timescales, multiple optimization objectives, multiple zones, multiple buildings, and building microgrids.

\end{document}